\documentclass[superscriptaddress, amsmath,amssymb, aps, twocolumn,longbibliography]{revtex4-2}
\usepackage[plainpages = false, pdfpagelabels, 
                 pdfpagelayout = useoutlines,
                 bookmarks,
                 bookmarksopen = true,
                 bookmarksnumbered = true,
                 breaklinks = true,
                 linktocpage,
                 colorlinks = true,  
                 linkcolor = blue,
                 urlcolor  = blue,
                 citecolor = blue,
                 anchorcolor = green,
                 hyperindex = true,
                 hyperfigures
                 ]{hyperref} 

\usepackage{enumerate}
\usepackage[caption=false,singlelinecheck=off]{subfig}
\usepackage{slashed}
\usepackage{braket}
\usepackage{subfloat}
\usepackage{bbold}
\usepackage{booktabs}
\usepackage{blindtext}
\usepackage{amsfonts}
\usepackage{bbm}
\usepackage{graphicx}
\usepackage{dcolumn}
\usepackage{bm}
\usepackage{amsmath, amssymb}
\usepackage[normalem]{ulem}
\newcommand{\NEW}[1]{{}{\textcolor{black}{ #1}}}
\renewcommand{\vec}[1]{{\boldsymbol #1}}
\newcommand{\short}[1]{{}{}}

\begin{document}
\title{\short{Emerging Majorana zero modes as a manifestation of edge-bulk  incompatibility} Manifestation of edge-bulk incompatibility in fractional quantum Hall platform}
\author{Jinhong Park}
\affiliation{Institute for Quantum Materials and Technologies, Karlsruhe Institute of Technology, 76021 Karlsruhe, Germany}
\affiliation{Institut für Theorie der Kondensierten Materie, Karlsruhe Institute of Technology, 76128 Karlsruhe, Germany}
\author{Yuval Gefen}
\affiliation{Department of Condensed Matter Physics, Weizmann Institute of Science, Rehovot 76100, Israel}

\date{\today}
\begin{abstract}
\NEW{The edges of a two-dimensional topological phase of matter serve as a platform underlying its low-energy dynamics. The topology of the bulk phase dictates the structure of the gapless modes. Proximitizing boundary modes to another boundary, may lead to gap opening at the edge. Subsequently, one may engineer different segments of the boundary with intrinsic incompatibility of gap-generating mechanisms ("edge-edge incompatibility"),   facilitating  the generation of topological excitations, e.g. Majorana zero modes (MZMs). Here we address the possibility of bulk-edge incompatibility, whereby the intrinsic bulk gap competes with a gap generated via boundary modes.  Specifically, we consider two $\nu = 2/3$ fractional quantum Hall phases whose shared boundary modes are gapped out via disorder-generated tunneling across the boundary.  A “neutral superconducting” phase, made up of neutral edge modes, is stabilized over a broad range of interaction parameters. This phase cannot coexist with the bulk gap. The resulting edge-bulk incompatibility gives rise to the emergence of MZMs (in the neutral sector). We propose an experimental setup to verify both the neutral superconductivity phase and the emergent MZMs.} 
\end{abstract}
\maketitle

\NEW {The nature of gapless modes at the boundary of two-dimensional topological phases of matter (prime examples thereof are fractional quantum Hall (FQH) phases), is tightly constrained by the topological quantum numbers of the bulk, giving rise to quantized values of various transport quantities, e.g., the electrical and thermal conductance~\cite{Banerjee2017, Banerjee2018}. This bulk-boundary correspondence allows to observe exotic properties of bulk quasiparticles via edge probes; an apt example is the observation of fractional braiding statistics of anyons~\cite{Nakamura2020,Nakamura2023,Kundu2023,Bartolomei2020,Rosenow2016,Han2016,Lee2019,Lee2022,Lee2023}. It may also predict the presence of upstream edge modes~\cite{Kane1994, Kane1995, Wang2013, Park2020}, which was confirmed in the measurement of heat transport~\cite{Banerjee2017,Venkatachalam2012, Altimiras2012, Melcer2022,Srivastav2022} and electrical noise~\cite{Bid2010, Gross2012, Kumar2022}.}

\begin{figure} [t]
\includegraphics[width=0.95\columnwidth]{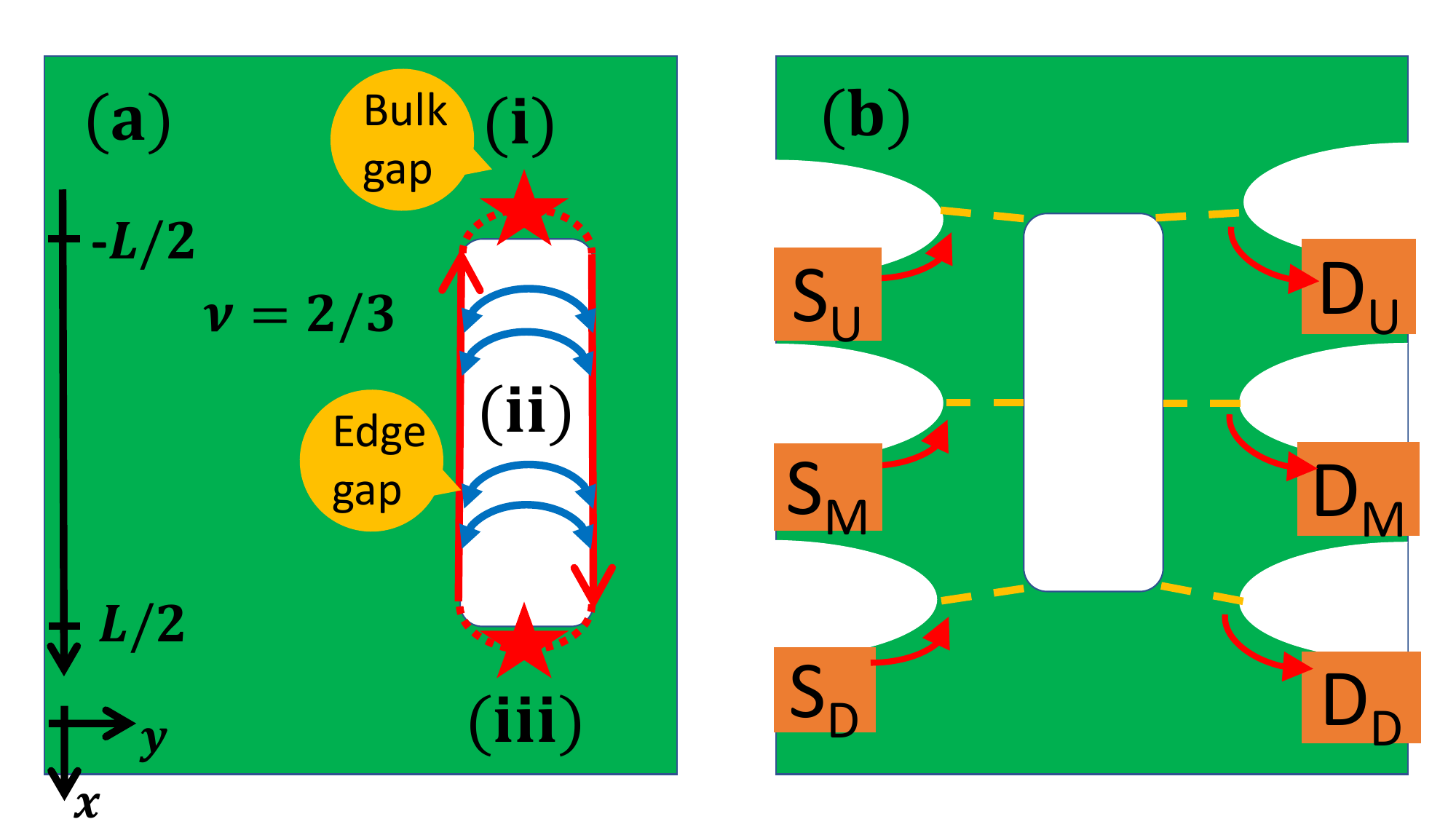} 
\caption{ {\bf Edge-bulk incompatibility,} {\bf (a)} A quantum anti-wire (QAW), supporting a $\nu = 0$ vacuum state (white), is surrounded by a $\nu = 2/3$ FQH liquid (green). The edge-mode gapping mechanism of the QAW competes with that of the $\nu = 2/3$ bulk phase. As function of $x$ ($y=0$) we move from (i) a bulk to and (ii) an edge to (iii) a bulk gap, Majorana zero modes (MZMs) (red starts) emerge at the incompatibility points: the (i)-(ii) boundary and the (ii)-(iii) boundary, marked in red stars. 
{\bf (b)} An array for a set of transport experiments for the detection of neutral-sector MZMs through transport experiments. Neutral quasi-particles (along with charged quasi-particles) are injected from the leads (labelled as S and D) through tunneling contacts (dashed yellow lines) to the QAW edge.  
}
\label{fig3:Antidotsetup}
\end{figure}

\NEW{Interestingly, it is possible to manipulate the boundary between two QH phases by alternatingly coupling segments of it to bulk superconductors (SC) and ferromagnets (FM), respectively~\cite{Nilson2008, Fu2009, Lindner2012, Clarke2013,Mong2014, Snizhko2018,Ebisu2017}. Disorder-induced tunneling across such a boundary may gap the edge spectrum; the gapping mechanisms of the SC-proximitized and the FM-proximitized segments are incompatible. This edge-edge incompatibility leads to the emergence of point-like topological excitations~\cite{Kitaev2003, Alicea2012, Leijnse2012, Beenakker2013} at the boundaries between segments. Specific protocols have been proposed~\cite{Nilson2008, Fu2009, Lindner2012, Clarke2013,Mong2014, Snizhko2018,Ebisu2017, Lutchyn2010, Oreg2010, Mourik2012, Fu2008, Zhang2014, Orth2015,Barkeshli2013, Barkeshli2014,Barkeshli2014arXiv} for the observation of MZMs or parafermions in such setups.}



\NEW{Here we bring forward a new type of incompatibility, that of bulk vs edge, that leads to the emergence of topological excitations. This incompatibility is underlain by the conflicting gapping mechanisms at the boundary and in the bulk. Notwithstanding the general applicability of this  notion, the specific case addressed here carries certain intriguing features:  while the  bulk gap is  a manifestation of FQH physics, the gap at the boundary is a consequence of an effective attractive interaction between neutral modes, stabilizing  a neutral superconducting phase. Proximitizing with extraneous superconductors is not required, thus avoiding the need for high resolution spatial tuning of boundary couplings.}

\NEW{Our geometry comprises an elongated quantum anti-dot (denoted as a quantum anti-wire (QAW)), inserted in a $\nu=2/3$ FQH bulk phase (cf. Fig.~\ref{fig3:Antidotsetup}(a)). The end-points of the QAW mark the separation between bulk and edge. Edge-bulk incompatibility gives rise to a topological mode at this separation point—here these are MZMs. We propose experimental manifestations for the observation of neutral superconductivity and MZMs, hence the emergence of edge-bulk incompatibility.}

\begin{figure} [t]
\includegraphics[width=\columnwidth]{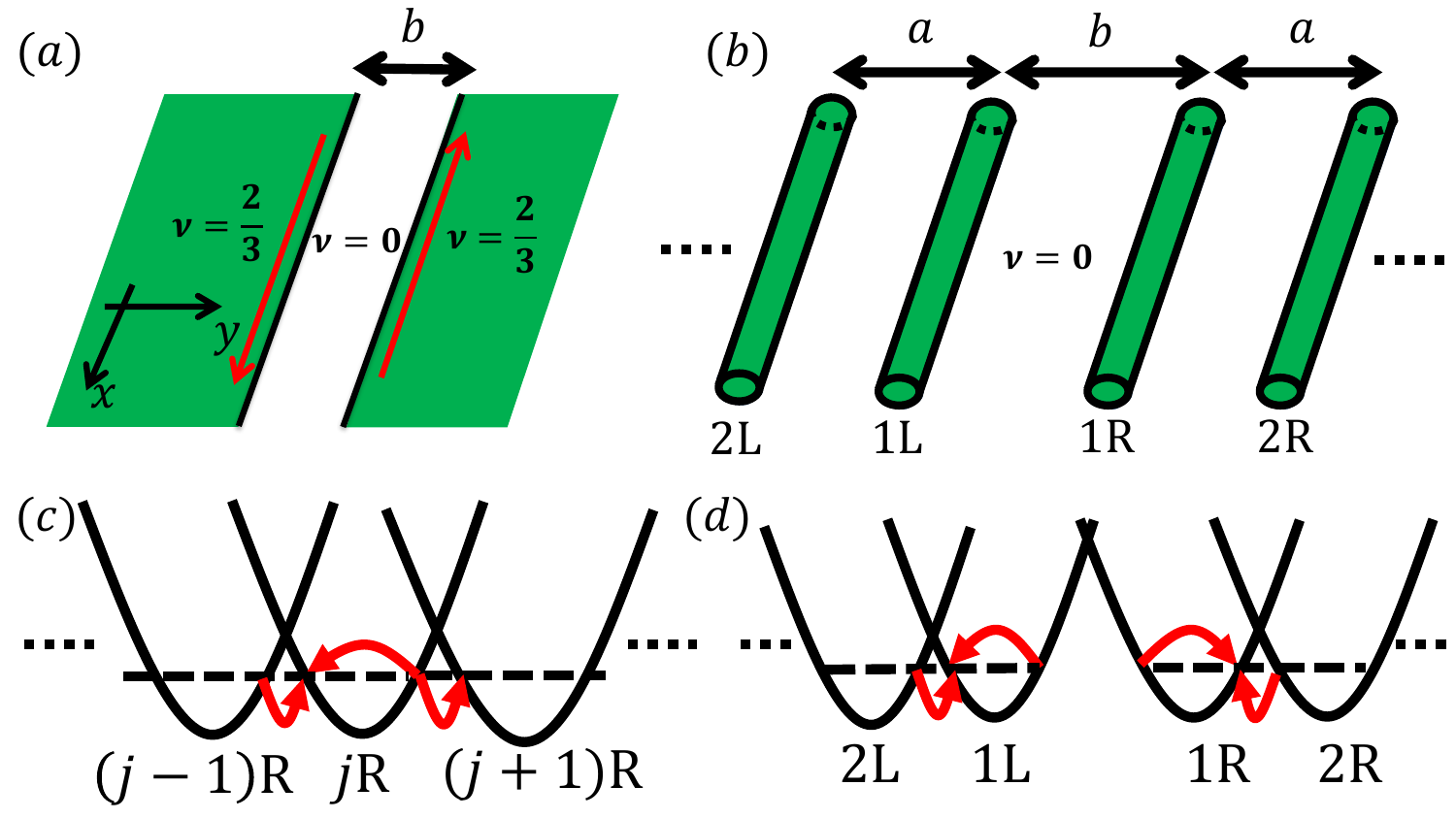} 
\caption{{\bf Model and setup.} {\bf (a)} Two $\nu = 2/3$ fractional quantum Hall (FQH) bulk phases (left L and right R) separated by a vacuum $\nu = 0$ state of width $b$. 
The boundary of the FQH bulks with the $\nu = 0$ state is described by two counter-propagating modes (red solid and blue dashed arrows). {\bf (b)}
Each of the FQH bulks comprises $2N$ wires ($j: 1 \rightarrow 2N $) that are separated by a distance $a$ from each other in $y$ direction.  
{\bf (c)} Each of the wires has a quadratic dispersion in $k_x$, centered at $|e| B a j/\hbar$ in the Laudau gauge $\vec{A} = (-By, 0, 0)$. 
The wires are gapped out by momentum conserving three-particle processes (cf.~Eqs.~\eqref{eq:threeparticleprocess} and \eqref{eq:bulkgap}), resulting in 
4 counterpropagating modes at the boundary. 
{\bf (d)} The boundary modes can be gapped out by the depicted process, resulting in {\it neutral superconductivity}. Disorder is required for this process since 
momentum is not conserved.
}
\label{Wireconstruction}
\end{figure}

\short{{\it A boundary between two bulk phases.---}}\NEW{{\it Bulk gap-forming mechanism.---}}As a first step, we consider a translationally invariant boundary (along the $x$ direction) between two semi-infinite $\nu = 2/3$ spin-polarized FQH bulk puddles (left L and right R),  depicted in Fig.~\ref{Wireconstruction}(a). The parallel edges of these two bulk puddles are positioned a distance $b$ apart. \short{To understand which microscopic process opens up a gap in the bulk puddles, we represent each puddle through a wire construction}

\NEW{To understand the gapping mechanism in the bulk puddles, we represent each puddle through a wire construction}~\cite{Kane2002, Teo2014, Meng2015, Meng2020, Sagi2014, Sagi2015, Kainaris2018}.
The anisotropic wire construction of a $\nu=2/3$ FQH phase consists of $2N$  wires (denoted as $j = 1, \cdots, 2N$) along the $x$ direction separated by a distance $a$ from each other (Fig.~\ref{Wireconstruction}(b)). We denote the fermionic creation operator as $\psi_{r, j, \eta}^{\dagger}$ for the propagating mode with chirality $\eta = \pm$ of wire $j$ in the bulk puddle $r = \textrm{L}, \textrm{R}$. The bulk of, e.g., puddle R, 
is gapped by a three-particle momentum conserving interaction~(Fig.~\ref{Wireconstruction}(c)),
\begin{align}  \label{eq:threeparticleprocess}
O_{\textrm{R}, j} (x) & = \psi_{R, j+1, -}^{\dagger} (\psi_{R, j, +})^2
 (\psi_{R, j, -}^{\dagger})^2  \psi_{R, j-1, +}
\nonumber \\ & \sim e^{ i \left (\tilde{\phi}_{\textrm{R},j+1}^{-} + 2 \tilde{\phi}_{\textrm{R},j}^{+}+ 2 
\tilde{\phi}_{\textrm{R},j}^{-}+ \tilde{\phi}_{\textrm{R}, j-1 }^{+} \right )}. 
\end{align}
The fermionic operator is bosonized as $\psi^{\dagger}_{R, j, \eta} (x) \sim e^{ - i \eta \tilde{\phi}_{R,j}^{\eta}(x)}$ with bosonic fields $\tilde{\phi}_{R,j}^{\eta}$. 
Upon considering a pair of wires of $j = 2k - 1$ and $j = 2k$ ($1 \leq k \leq N$) and defining, for every $\eta$ and $k$, two chiral bosonic fields 
\begin{align}
\phi_{\textrm{R}, k, 1}^{\eta} &\equiv \tilde{\phi}_{\textrm{R}, 2k-1}^{\eta} + (1 + \eta) ( \tilde{\phi}_{\textrm{R}, 2k}^{+}+  \tilde{\phi}_{\textrm{R}, 2k}^{-} ) \nonumber \\  \phi_{\textrm{R}, k, 2}^{\eta} &\equiv \tilde{\phi}_{\textrm{R}, 2k}^{\eta} + (1-\eta)( \tilde{\phi}_{\textrm{R}, 2k-1}^{+}+ \tilde{\phi}_{\textrm{R}, 2k-1}^{-} ) , 
\end{align}
the gap forming interaction \eqref{eq:threeparticleprocess} becomes 
\begin{align} \label{eq:bulkgap}
    O_{\textrm{R}, 2k + 1} &\sim  e^{i ( \phi_{\textrm{R}, k, 2}^{+} + 
    \phi_{\textrm{R}, k+1, 2}^{-})} \nonumber \\
    &= e^{3 i  (\phi_{\textrm{c},\textrm{R}, k}^{+} + \phi_{\textrm{c},\textrm{R}, k+1}^{-})} e^{- i  (\phi_{\textrm{n},\textrm{R}, k}^{-} + \phi_{\textrm{n},\textrm{R}, k+1}^{+})},
    \nonumber \\ 
    O_{\textrm{R}, 2k} &\sim  e^{i ( \phi_{\textrm{R}, k, 1}^{+} + 
    \phi_{\textrm{R}, k+1, 1}^{-})} \nonumber \\
    &=e^{3 i  (\phi_{\textrm{c},\textrm{R}, k}^{+} + \phi_{\textrm{c},\textrm{R}, k+1}^{-})}
    e^{i  (\phi_{\textrm{n},\textrm{R}, k}^{-}+ \phi_{\textrm{n},\textrm{R}, k+1}^{+})}.  
\end{align}
Here we further defined charge modes $\phi_{c, R, k}^{\eta} \equiv \frac{1}{6} (\phi_{\textrm{R}, k, 1}^{\eta} + \phi_{\textrm{R}, k, 2}^{\eta})$ and neutral modes $\phi_{n, R, k}^{\eta}\equiv \frac{1}{2} (\phi_{\textrm{R}, k, 1}^{-\eta} - \phi_{\textrm{R}, k, 2}^{-\eta})$. 
\NEW{Below we show that these gap-forming operators \eqref{eq:bulkgap} may be incompatible with those in the boundary.}

When both operators defined in Eqs.~\eqref{eq:bulkgap} are renormalization-group (RG) relevant, the spectrum of each bosonic charge operator (e.g., $\phi_{c, \textrm{R}, k-1}^{+}$ and  $\phi_{c, \textrm{R}, k}^{-}$) and each bosonic neutral operator (e.g., $\phi_{n, \textrm{R}, k-1}^{-}$ and $\phi_{n, \textrm{R}, k}^{+}$) in the bulk is separately gapped. At the same time, boundary modes $\phi_{c}^{-} \equiv \phi_{c, \textrm{R}, 1}^{-}$ and 
$\phi_{n}^{+}  \equiv \phi_{n, \textrm{R},1}^{-}$ of puddle $R$, as well as $\phi_{c}^{+} \equiv \phi_{c, \textrm{L}, 1}^{+}$ and  $\phi_{n}^{-} \equiv \phi_{n, \textrm{L}, 1}^{-}$ of puddle $L$ remain gapless. Then, the boundary of each puddle consists of two counter-propagating modes, a charge and a neutral mode (cf. Fig.~\ref{Wireconstruction}(a)). 
The boundary modes satisfy the following commutation relation
\begin{align} \label{eq:commutationrelation}
[\phi_{\ell}^{\eta}(x), \phi_{\ell'}^{\eta'} (x')] = i \pi  \eta (K^{-1}_{0})_{\ell \ell'}\delta_{\eta \eta'} \text{sgn} (x - x')\,,
\end{align}
with $\ell, \ell' = n, c$ 
and $K_0=\begin{pmatrix}
   2 & 0 \\
  0 & 6 \\
   \end{pmatrix}$. 
Notwithstanding the unrealistic anisotropy present in the wire construction approach, this approach represents essential features of edge structure, 
importantly the emergence of counter-propagating boundary modes of the
$\nu = 2/3$ phase. Such counter-propagating modes are confirmed by the observation of upstream modes through thermal transport and noise measurements~\cite{Banerjee2017, Melcer2022, Kumar2022, Srivastav2022} in remarkable agreement with the theoretical predictions of Refs.~\cite{Kane1997, Protopopov2017, Park2019, Spanslatt2019}.

\short{{\it Coupling of the boundary modes.---}}\NEW{{\it Boundary gap-forming mechanism.---}}We now consider electron tunneling-induced as well as interaction-induced intermode coupling among the boundary modes $\phi_c^{\pm}$ and $\phi_n^{\pm}$.
The low-energy action for the boundary modes can be split into two parts, $S = S_0 + S_{\rm{inter}}$. 
$S_0$ comprises contributions due to bare non-interacting dynamics, as well as density-density interaction term. In the basis $\vec{\phi} = (\phi_{n}^{+}, \phi_{c}^{+}, \phi_{n}^{-},\phi_{c}^{-})^T$, it is written as 
\begin{align} \label{eq:forward}
S_{0} = & - \frac{1}{4\pi} \int dx dt \left[\left (\partial_x \vec{\phi}^T \right )\cdot ( K \cdot \partial_t \vec{\phi} + V \cdot \partial_x \vec{\phi})  \right]\,.
\end{align} 
Here $K=\begin{pmatrix}
   K_0 & 0 \\
   0 & -K_0 \\
    \end{pmatrix}$ and the interaction matrix $V$ is
\begin{align}
V =\begin{pmatrix} \label{interactionmatrix}
v_n & v_{cn}^{++} & g_n & v_{cn}^{+-} \\ 
v_{cn}^{++} & v_c & v_{cn}^{-+} & g_c \\ 
g_n & v_{cn}^{-+} & v_n & v_{cn}^{--} \\ 
v_{cn}^{+-} & g_c  & v_{cn}^{--} & v_c
\end{pmatrix},
\end{align}
with real-valued parameters $v_c$, $v_n$, $g_c$, $g_n$, $v_{cn}^{\eta \eta'}$; $\eta, \eta' = \pm$. 
In order to determine the possible phases of the boundary modes we consider the following Hamiltonian for the wires,
\begin{align} \label{eq:wireinteraction}
   & H_{\rm{wire}} =  \int dx \big (\sum_{r=R,L} \big ( -i \hbar v_F \sum_{j=1}^{2N} \sum_{\eta = \pm} \eta \psi_{r, j, \eta}^{\dagger} \partial_x \psi_{r, j, \eta}
    \nonumber \\ 
      &+ \frac{U}{2}\sum_{j=1}^{2N} \rho_{r, j}^2  
   + V \sum_{j=1}^{2N-1} \rho_{r, j}\rho_{r, j+1}\big )  + V' \rho_{L, 1} \rho_{R,1} \big), 
\end{align}
with the long-wavelength electron density $\rho_{r, j} (x) \equiv \sum_{\eta = \pm } \psi_{r,j,\eta}^{\dagger} (x) \psi_{r,j,\eta} (x)$.
Here $v_F$ is bare Fermi velocity of the wires, $U$ the intra-wire interaction, $V$ the nearest neighbor intra-puddle interaction, and $V'$ is the nearest neighbor inter-puddle interaction. Using this microscopic Hamiltonian and omitting the bulk gapped modes, we obtain Eq.~\eqref{eq:forward} with the parameters $v_c = 14 v_F + (U+V)/\pi$,  $v_n = 6 v_F  + (U-V)/\pi$, $v_{cn}^{\pm \mp} = \pm 8v_F$, $g_c = V'/(2 \pi)$, $g_n = - V'/(2 \pi)$, and $v_{cn}^{\pm \pm} = \pm V'/(2 \pi)$. Note that for a repulsive interaction $V'>0$, the interaction $g_n$ between $\partial_x\phi_n^+$ and $\partial_x\phi_n^-$ is negative, i.e., {\it attractive}. If this effective
attractive interaction is strong, it stabilizes the neutral superconductivity
phase as will be shown below.

\begin{figure} [t]
\includegraphics[width=0.95\columnwidth]{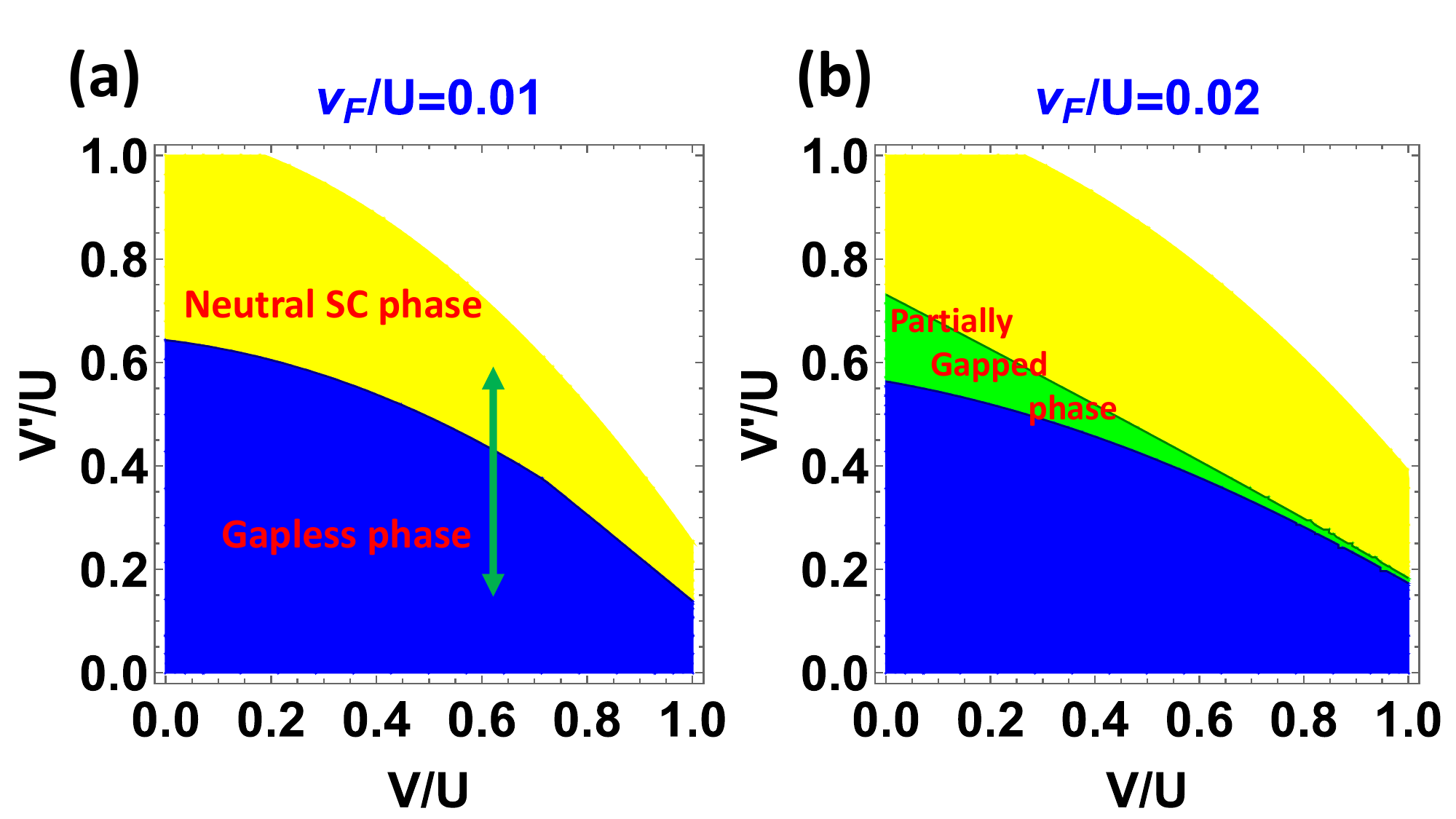} 
\caption{{\bf Phase diagrams for the neutral sector associated with the boundary} 
in the parameter space ($V/U$, $V'/U$) with bare velocity {\bf (a)} $v_F/ U =0.01$ and {\bf (b)} $v_F/ U =0.02$, see Eq.~\eqref{eq:wireinteraction} for the parameters. 
Three distinct phases appear at zero temperature: neutral superconducting phase (in yellow), partially gapped phase (green), and gapless phase (blue). While $\hat{O}_{\text{nsc}}^{\pm}$ is the most renormalization group (RG) relevant, the gapless phase is characterized by the intra-bulk puddle operators $\hat{O}_{\text{intra}}^{\pm}$ being the most RG relevant, see Eq.~\eqref{eq:scatteringterm}. \NEW{The green line in (a) depicts schematically the variation of the relevant  parameters with the  gate voltage or the magnetic field.}
In (b), there emerges a narrow region where two among 4 boundary modes are gapped, leaving the two other modes gapless. Those three phases are characterized by distinct central charges; $c=2, 1,0$ in the blue, green, and yellow regions, respectively. In the white region, the boundary modes are unstable since the $V$-matrix \eqref{interactionmatrix} is not positive-definite. 
}
\label{PhaseDiagram}
\end{figure}

$S_{\rm inter}$ represents processes 
to describe tunneling in and out of the boundary modes that change the density of individual modes, 
\begin{align} \label{eq:intermode-scattering}
   S_{\rm{inter}} = \int dx dt \left ( \hat{O}_{\textrm{nsc}}^{\pm} +\hat{O}_{\textrm{inter}}^{\pm} +  \hat{O}_{\textrm{intra}}^{\pm} + \text{h.c.} \right ).
\end{align}
Here, $\hat{O}_{\textrm{nsc}}^{\pm}$ represents {\it neutral superconductivity} operators, $\hat{O}_{\rm{inter}}^{\pm}$ inter-puddle electron-tunneling operators, and $\hat{O}_{\rm{intra}}^{\pm}$ intra-puddle electron-tunneling operators, 
\begin{align} \label{eq:scatteringterm}
\hat{O}_{\textrm{nsc}}^{\pm} (x)&= \xi_{\textrm{nsc}}^{\pm} e^{\pm i (\phi_n^{+} - \phi_{n}^{-})} e^{3 i  (\phi_c^{+} + \phi_{c}^{-})}\,, \nonumber \\ 
\hat{O}_{\textrm{inter}}^{\pm} (x)&= \xi_{\textrm{inter}}^{\pm} e^{\pm i (\phi_n^{+} + \phi_{n}^{-})} e^{3 i  (\phi_c^{+} + \phi_{c}^{-})}\,,\nonumber \\ 
\hat{O}_{\textrm{intra}}^{\pm} (x)&= \xi_{\textrm{intra}}^{\pm} e^{2 i  \phi_n^{\pm}}\,.
\end{align}
Whereas $\hat{O}_{\textrm{intra}}^{\pm}$ couple the boundary modes in the same puddle, $\hat{O}_{\textrm{nsc}}^{\pm}$ and $\hat{O}_{\textrm{inter}}^{\pm}$ are associated with electron tunneling between different puddles. One concludes from Eq.~\eqref{eq:commutationrelation} that $\hat{O}_{\textrm{nsc}}^{\pm}$ shifts the neutral density $\hat{\rho}_n \equiv \partial_x (\phi_n^+ + \phi_{n}^-) / \pi$ : 
\begin{align}
    \hat{O}_{\textrm{nsc}}^{\pm} (x') \hat{\rho}_n (x) (\hat{O}_{\textrm{nsc}}^{\pm})^\dagger (x')
    = \hat{\rho}_n (x) \pm 2\delta (x - x')\,,
\end{align}
\NEW{suggesting that $\hat{O}_{\textrm{nsc}}^{\pm}$ is associated with the neutral superconductivity (the factor $2$ indicates pairing). The operator  $\hat{O}_{\textrm{nsc}}^{\pm}$, inducing a superconducting gap in the neutral sector,  differs from Eqs.~\eqref{eq:bulkgap}, which is responsible for gap generation in the bulk. This underlies the bulk-edge incompatibility.} 
Interestingly, $\hat{O}_{\textrm{nsc}}^{\pm}$, Eq.~\eqref{eq:scatteringterm}, also involves the charge sector. It should be then contrasted with the neutral superconductivity terms investigated in Ref.~\cite{Jukka2022}, where the latter do not involve the charge sector.



The terms in Eq.~\eqref{eq:intermode-scattering} are a direct consequence of disorder along the boundary (see Fig.~\ref{Wireconstruction}d and supplements~\cite{Supple}). The disorder-induced terms are 
assumed to follow a Gaussian distribution, characterizing the disorder strength $W_{\ell}^{\pm}$ by
\begin{align} \label{eq;disorderstrength}
\langle \xi^{\pm}_{\ell} (x) (\xi^{\pm}_{\ell'})^* (x') \rangle_{\text{dis}} = W_{\ell}^{\pm} \delta_{\ell \ell'} \delta (x - x')
\end{align}
for $\ell=\textrm{nsc}, \textrm{inter}, \textrm{intra}$. For weak disorder, $W_{\ell}^{\pm}$ renormalize according to the RG equations
$d W_{\ell}^{\pm}/d \ln \mathcal{L} = (3 - 2 \Delta_{\ell}^{\pm} ) W_{\ell}^{\pm}$
with the corresponding scaling dimension $\Delta_{\ell}^{\pm}$ and the running length $\mathcal{L}$~\cite{Giamarchi1988}.

{\it Phase diagram.---}
The zero-temperature phase diagram for the boundary modes
is obtained by identifying the most relevant (smallest scaling dimension) operator of Eq.~\eqref{eq:scatteringterm} (w.r.t. the action \eqref{eq:forward} with parameters deduced from Eq.~\eqref{eq:wireinteraction}), see SM~\cite{Supple}.
The phase diagrams depicted in Fig.~\ref{PhaseDiagram} are mainly governed by two 
distinct phases in the ($V/U$, $V'/U$) plane: a {\it neutral superconducting phase} (yellow) and a {\it gapless phase} (blue).
\NEW{While $\hat{O}_{\text{nsc}}^{\pm}$ are the most RG-relevant in the neutral superconducting phase, the gapless phase is characterized by $\hat{O}_{\text{intra}}^{\pm}$ being the most RG-relevant. According to the Haldane criterion~\cite{Haldane1995}, the  latter cannot open a gap.  With sufficiently large $V, V'$ and on-site interaction $U \gg v_F$,
the neutral superconductivity phase becomes stable. In such a neutral superconductivity phase, generating inter-puddle neutral "cooper pairs" is energetically favorable, leading to gapped neutral excitations. Furthermore, forming neural superconductivity phase at the boundary generates a conflicting gap with the bulk gap.} 

\short{{\it Neutral MZMs in a quantum-wire setup.---}}\NEW{{\it Manifestations of edge-bulk incompatibility.---}} We \short{now }consider an elongated quantum anti-dot  (QAW) geometry of length $L$ (Fig.~\ref{fig3:Antidotsetup}(a)) that breaks the translational invariance along the $x$ direction. This QAW can be formed by locally depleting the electron density. Such a QAW represents an artificial "defect" that can host neutral MZM excitations. The system comprises three  domains: (i) $x<-L/2$, (ii) $|x|<L/2$, 
(iii) $x>L/2$. Following the $y = 0$ dashed yellow line, we move from a $\nu=2/3$ bulk phase (domain (i)) to a vacuum strip separating two such $\nu= 2/3$ phases (domain (ii)), and back to a single bulk phase (domain (iii)). In the corresponding wire-construction picture, domain (i) and (iii) are characterized by the bulk-gap forming interaction, Eq.~\eqref{eq:bulkgap}, 
while domain (ii) is tuned to exhibit the neutral superconductivity phase (yellow region in Fig.~\ref{PhaseDiagram}), having $\hat{O}^\pm_{\rm nsc}$ the most relevant operators. It follows that the Hamiltonian along the $y = 0$ line is given by
\begin{align} 
\label{eq:qawHam}
 &H_{\text{QAW}} = \sum_{\eta = \pm}\int \frac{dx}{4\pi}  
\Big ( v_c (\partial_x \phi_c^{\eta})^2 + v_n (\partial_x \phi_n^{\eta})^2 \Big ) \nonumber 
\\
&-  2t_{\text{b}}\int_{\text{(i)} \cup \text{(iii)} } dx  
\cos(\phi_{n}^{+} + \phi_{n}^{-} ) \cos (3 (\phi_{c}^{+} + \phi_{c}^{-}))
\nonumber \\ 
&- \sum_{p = \pm} \int_{\text{(ii)}} dx |\xi_{\text{nsc}}^{p}|
\cos(\phi_{n}^{+} - \phi_{n}^{-} +3 p\ (\phi_{c}^{+} + \phi_{c}^{-})  + \zeta^{p})\,.
\end{align}
Note that $\xi_{\text{nsc}}^{\pm}(x) = |\xi_{\text{nsc}}^{\pm}(x)| e^{i  \zeta^{\pm} (x)}$ in domain (ii) is induced by weak disorder (hence it is fluctuating with $x$) while $t_{\text{b}}$ in domain (i) and (iii) is constant. If $t_{\text{b}}$ is sufficiently strong, the charge and neutral modes in domain (i) and (iii)
are separately pinned to satisfy 
\begin{align}
 \theta_c \equiv \phi_c^+ + \phi_c^- &= \frac{\pi \hat{m}_c}{3}\,, \nonumber \\
 \theta_n \equiv \phi_n^+ + \phi_n^- &= \pi \hat{m}_c + 2 \pi \hat{m}_n \,,
\end{align}
with integer valued operators $\hat{m}_c$ and $\hat{m}_n$. Those pinning gap out both charge and neutral sectors. 
For strong $|\xi^{\pm}_{\rm nsc}|$ in domain (ii), 
the third term in Eq.~\eqref{eq:qawHam} would favor finding a minimum of the cosine term for all values of $x$ such that
\begin{align} \label{eq:pinningregimeii}
\theta_c \equiv \phi_c^+ + \phi_c^- &= \frac{\pi \hat{m}_c}{3} + \frac{\zeta^+ - \zeta^-}{6}\,, \nonumber \\
\phi_n \equiv \phi_n^+ - \phi_n^- &= \pi \hat{m}_c + 2 \pi \hat{l}_n +  \frac{\zeta^+ + \zeta^-}{2}\,,
\end{align}
where $\hat{l}_n $ is another integer operator. 
But, such local random quenching of $\theta_c$ and $\phi_n$ competes against  minimizing 
the kinetic energy (the first term in Eq.~\eqref{eq:qawHam}). 
This results in 
$\theta_c$ and $\phi_n$  varying slowly over the localization length $\ell$,
which scales as 
$\ell \propto \text{min}_{p = \pm} \left (1/((W_{\textrm{\rm{nsc}}}^{p,0})^{1/(3-2\Delta_{\rm{nsc}}^{p})}) \right)$.
Here $W_{\textrm{\rm{nsc}}}^{\pm,0}$ are the bare disorder strengths for 
$\xi_{\rm{nsc}}^{\pm}$.

At the domain walls ($|x| = L/2$), the two non-commuting operators, $\hat{l}_n$ and $\hat{m}_n$, compete with each other, and hence the two operators cannot be pinned simultaneously, leading to zero-energy neutral excitations at the domain walls. The elementary neutral excitations are given by, up to the trivial phase factor,  
\begin{align} \label{eq:basicneutralexc}
    \hat{\alpha}_n^{\eta} \equiv e^{i \eta \phi_n^\eta} = e^{i \eta  \pi (\hat{m}_c + \hat{m}_n + \hat{l}_n)}\,.
\end{align}
These excitations satisfy $(\hat{\alpha}_n^{\eta})^2 = 1$, a basic requirement from MZMs.
By contrast, generation of the charge excitations involves energy $\Delta_c \sim \hbar v_c/\ell$. 

\begin{figure} [t]
\includegraphics[width=\columnwidth]{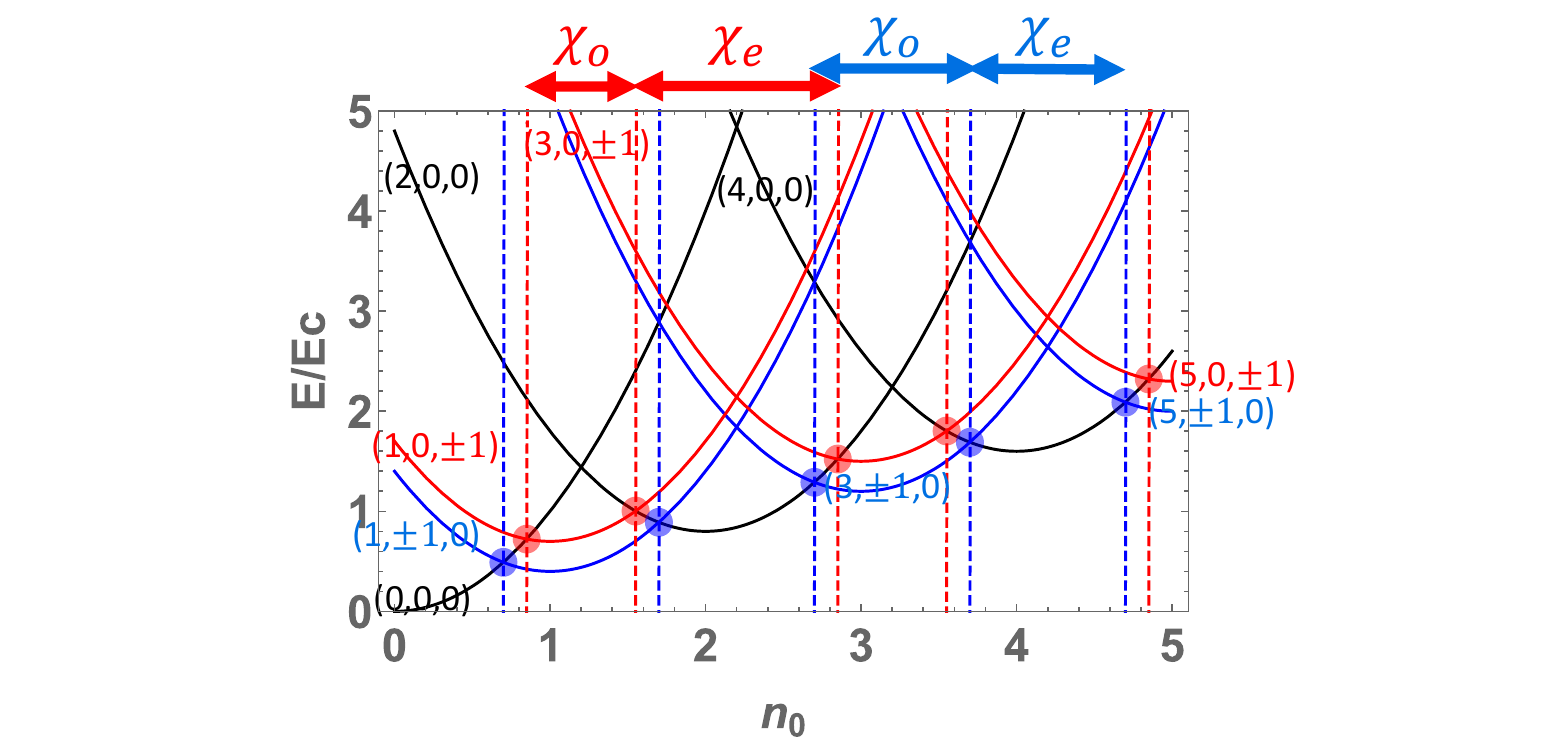} 
\caption{{\bf Energy spectra} of the QAW setup, $E(N_c, N_n^{0}, N_n^M)$ versus $n_0$, which is controlled by gates underneath the QAW. Energy spectra are characterized by three quantum numbers ($N_c, N_n^{0}, N_n^M$), see Eq.~\eqref{eq:energyspectra}. While $N_c$ ($N_n \equiv N_n^{0}+ N_n^M$) is the number of charge (neutral) excitations in the QAW, $N_n^{M}$ and $N_n^{0}$ represent
the occupation of neutral MZMs and the neutral single-particle states above a superconducting gap, respectively. 
Black, blue, red parabolas indicate the $N_n = 0$, $N_n^M = \pm 1$, $N_n^0 = \pm 1$ states, respectively. 
When source S$_\text{M}$ is baised while the current is measured in drain D$_\text{M}$ in Fig.~\ref{fig3:Antidotsetup}(b), the (differential) conductance has a zero-bias peak at red circles, i.e., $E (N_c, 0,0) = E(N_c+1, \pm 1,0)$. Due to the neutral superconducting gap, the spacing of the peaks has an even-odd effect with $\chi_e \neq \chi_o$. In contrast, when source S$_\text{U}$ is baised while the current is measured either in D$_\text{D}$ (non-local measurement) or D$_\text{U}$ (local), the peaks occur at blue circles, $E (N_c, 0,0) = E(N_c+1, 0, \pm 1)$.
The occupation of the MZMs does not cost energy so that the peaks are equally spaced, $\chi_e = \chi_o$. Parameters: $\Delta_c = 0.4 E_c$ and $\Delta_n = 0.3 E_c$.}
\label{fig4:QAWcond}
\end{figure}

\NEW{To experimentally tune domain (ii) (cf. Fig.~\ref{fig3:Antidotsetup}) to the neutral superconductivity phase, one slightly varies the magnetic field and the gate voltage underneath the QAW, while keeping domain (ii) depleted. Those tunings allow to effectively change the distance between edge modes across the QAW and thus control the magnitude of $V'$ along the green lines in Fig.~\ref{PhaseDiagram}a. This amounts to inducing a phase transition from a gapless phase (blue) to a neutral superconductivity phase (yellow), see e.g., Ref.~\cite{Cohen2019} for an experiment to control couplings between edge modes employing the above knobs. }

\NEW{{\it Experimental signatures.---}}To detect neutral superconductivity and the emergence of MZMs in the QAW setup, we propose a set of transport experiments, see Fig.~\ref{fig3:Antidotsetup}(b). The energy of the QAW is characterized by three quantum numbers ($N_c$, $N_n^0$, $N_n^M$); $N_c$ is the number of charge excitations, $N_n^M = 0, 1$--- the occupation of neutral MZMs, $N_n^0$--- the number of neutral single-particle excitations above superconducting gap $\Delta_n \sim \hbar v_n/\ell$. The energy reads
\begin{align} \label{eq:energyspectra}
E (N_c, N_n^0, N_n^M) &= E_c (N_c - n_0 )^2  + \Delta_c N_c + \Delta_n \text{mod} (N_n^0, 2),
 \end{align}
where $n_0$ is a parameter effectively tuned by a back gate controlling the QAW. 
While the first term is the charging energy of the QAW that arises from the long-range Coulomb interaction, the second term corresponds to a single-particle gap $\Delta_c$. The neutral superconductivity in the QAW renders the third term in Eq.~\eqref{eq:energyspectra} dependent on the parity of $N_n = N_n^0+N_n^M$, i.e., even or odd occupations. This QAW is connected by tunneling contacts which facilitate tunneling of neutral and charged quasiparticles from the leads (labelled as $\text{S}$ and $\text{D}$), cf.~Fig.~\ref{fig3:Antidotsetup}(b). Importantly, such a tunneling process changes $N_c$ and $N_n$, keeping
$\delta N_c + \delta N_n$ is even~\cite{Naud2000, Supple}. 

Depending on the measurement scheme in Fig.~\ref{fig3:Antidotsetup}(b), the conductance peaks exhibit distinct even-odd behaviors as shown in Fig.~\ref{fig4:QAWcond}. When source S$_\text{M}$ is baised and the current is measured in drain D$_\text{M}$, the neutral superconductivity in the QAW leads to a significant even-odd behavior in the conductance peaks, i.e., $\chi_e \neq \chi_o$ (red dots in Fig.~\ref{fig4:QAWcond}). Since charge superconductivity is expected strongly suppressed in this system, this even-odd effect is an manifestation of neutral superconductivity. Such an even-odd effect exists provided that $E_c > \Delta_n$ and $L \gg \ell$. 
By contrast, when S$_\text{U}$ is baised while the current is measured in D$_\text{D}$ or D$_\text{U}$, one adds a neutral MZM at the end point of the QAW without energy cost. Thus the even-odd effect vanishes with equally spaced peaks, $\chi_e = \chi_o$ (blue dots).

\NEW{{\it Summary and outlook.---}Our introduction of the notion of edge-bulk incompatibility is followed by a detailed analysis of a specific case: that of the $\nu = 2/3$ FQH phase. The specifics of this case reveal the emergence of a stable neutral-sector superconducting phase, and the subsequent emergence of MZMs. These features can be tested through charge transport measurements. Generalizations of our theory include extension of the present analysis to other particle-hole conjugate FQH states with multiple neutral modes~\cite{Kane1995, Wang2013}. These may exhibit unusual noise features (auto- and cross-correlations). An intriguing direction would be to replace the QAW by an elongated island of a non-trivial topological phase (e.g., an island of a FQH phase), allowing us to manipulate the boundary, hence the edge-bulk incompatibility.}



\begin{acknowledgments}
We thank Alexander D. Mirlin and Tobias Meng for fruitful discussions. We acknowledge support by the DFG Grant MI 658/10-2 and by the German-Israeli Foundation Grant I-1505-303.10/2019.
Y.G. further acknowledges support by the Helmholtz International Fellow Award, by the US-Israel BSF Grant 714789, 
by CRC 183 (project C01), and by the Minerva Foundation. Y.G. is the incumbent of InfoSys Chair (at IISc).    
\end{acknowledgments}

\bibliographystyle{apsrev4-1-titles}
\bibliography{NeutralMZMarxiv}


\appendix

\onecolumngrid

\global\long\def\thesection{S\Alph{section}}
\global\long\def\thesubsection{\Roman{subsection}}
\setcounter{equation}{0}
\setcounter{figure}{0}
\setcounter{table}{0}
\setcounter{page}{1}
\renewcommand{\theequation}{S\arabic{equation}}
\renewcommand{\thefigure}{S\arabic{figure}}
\renewcommand{\bibnumfmt}[1]{[S#1]}

\bigskip
\begin{center}
\large{\bf Supplemental Material for "Manifestation of edge-bulk incompatibility in fractional quantum Hall platform"\\}
\end{center}
\begin{center}
Jinhong Park$^{1}$  and Yuval Gefen,$^{2}$
\\
{\it $^{1}$Institute for Theoretical Physics, University of Cologne, Z\"{u}lpicher Str. 77, 50937 K\"{o}ln, Germany \\$^{2}$Department of Condensed Matter Physics, Weizmann Institute of Science, Rehovot 76100, Israel} \\
(Dated: \today)
\end{center}

\section{Wire construction} \label{supsec:wireconstruction}
In this section, we will discuss a wire-construction approach for $\nu = 2/3$ quantum Hall (QH) puddles developed in Refs.~\cite{Kane2002,Teo2014}, and show that the boundary of each puddle exhibits two counter-propagating modes. This edge picture is largely consistent with the observation of upstream modes from a shot noise measurement and thermal transport~\cite{Banerjee2017, Melcer2022, Kumar2022, Srivastav2022}. 


We consider two semi-infinite spin-polarized $\nu = 2/3$ quantum Hall puddles, left (L) and right (R), 
separated by the $\nu = 0$ vacuum state with width $b$, see Fig.~2(a) in the main text. 
Each of the puddles consists of $2N$ wires along the $x$ direction, being distance $a$ apart from each other (see Fig.~2(b) in the main text). The wires have quadratic dispersion in $k_x$, given by the Hamiltonian in the Landau gauge $\vec{A} = (-By, 0, 0)$,
\begin{align}
    H_0 (k_x) = \sum_{r=\pm 1}\sum_{j = 1}^{2N} \frac{1}{2m} \big (\hbar k_x -r |e| B (\frac{b}{2} + a (j-1)) \big)^2\,,
\end{align}
with $r= \rm{R}/\rm{L} = \pm 1$. 
As shown in Fig.~2(c) in the main text, the parabola for the dispersion of each individual wire is centered at $k_x = \frac{r |e| B}{\hbar} (\frac{b}{2} + a (j-1))$. In the vicinity of the Fermi energy (the horizontal dashed line in Fig.~2(c)), the low-energy physics is captured by the linearized Hamiltonian (in the second quantization form)
\begin{align} \label{supeq:linearizedHamiltonian}
    H_0 =  \sum_{\eta = \pm 1} \sum_{r=\pm 1}\sum_{j = 1}^{2N}  \int dx   \psi_{r, j, \eta}^{\dagger} \hbar v_F \left ( (-i \eta \partial_x - k_F ) - \frac{r |e| B \eta}{\hbar} (\frac{b}{2} + a (j-1)) \right)\psi_{r, j, \eta }. 
\end{align}
Each parabola is described by two fermionic fields $\psi_{r, j, \eta = \pm 1}$, propagating along the opposite direction to each other.
$v_F$ is the bare Fermi velocity of the wires and the Fermi wavevector $k_F$ is related with the filling factor $\nu$ by $k_F = |e| a B \nu /(2 \hbar)$. A gauge transformation of 
\begin{align} \label{supeq:gaugetransformation}
\psi_{r, j, \eta} (x) \rightarrow  \psi_{r, j, \eta} (x)
e^{i \big (\eta k_F + \frac{r |e| B}{\hbar} (\frac{b}{2} + a (j-1)) \big ) x }
\end{align}
allows us to gauge away the $k_F$- and wire-dependent terms in Eq.~\eqref{supeq:linearizedHamiltonian} as 
\begin{align} \label{supeq:aftergaugetransformation}
      H_0 \rightarrow H_0=  \sum_{\eta = \pm 1} \sum_{r=\pm 1}\sum_{j = 1}^{2N}  \int dx   \psi_{r, j, \eta }^{\dagger}  \left ( -i \hbar v_F \eta \partial_x \right)\psi_{r, j, \eta }\,.
\end{align}

We next bosonize $\psi_{r, j, \eta}$ by introducing chiral bosonic fields $\tilde{\phi}_{r,j}^{\eta}$ that obey the commutation relation 
\begin{align} \label{supeq:commutationrelationoriginal}
[\partial_x \tilde{\phi}_{r,j}^{\eta} (x), \tilde{\phi}_{r',j'}^{\eta'} (x') ] = 2 i \pi \eta  \delta_{\eta \eta'} \delta_{r r'}  \delta_{j j'} \delta(x-x').
\end{align}
The fermionic fields are written in terms of $\tilde{\phi}_{r,j}^{\eta}$ as 
\begin{align} \label{supeq:fermionicoperators}
\psi_{r, j, \eta} (x) =  \frac{1}{\sqrt{2 \pi \epsilon}} e^{i \eta \tilde{\phi}_{r,j}^{\eta} (x)} e^{i \big (\eta k_F + \frac{r |e| B}{\hbar} (\frac{b}{2} + a (j-1)) \big ) x }, 
\end{align}
with the inclusion of the effect of the gauge transformation of Eq.~\eqref{supeq:gaugetransformation}. 
Here $\epsilon$ is a short-distance cutoff. 
In this bosonization description, the density fluctuations $\rho_{r, j, \eta} (x) \equiv \psi_{r, j, \eta}^{\dagger} (x) \psi_{r, j, \eta} (x) $ is given by 
\begin{align}
\rho_{r, j, \eta} (x) = \frac{1}{2\pi} \partial_x \tilde{\phi}_{r,j}^{\eta}. 
\end{align}

So far, the entire system is gapless. A wealth of correlated phase is realized by coupling the wires in different ways to open a gap~\cite{Kane2002,Teo2014}. In particular, the bulk gap of the $\nu = 2/3$ quantum Hall states is developed  by introducing 
three-particle involving processes~(cf. Fig.~2(c) in the main text), described by the following operators
\begin{align} \label{sup:threeparticleprocess}
O_{\textrm{R}, j} (x) &= \psi_{\textrm{R}, j+1, -}^{\dagger} \left ( \psi_{\textrm{R}, j, +}\right)^2 \left ( \psi_{\textrm{R}, j, -}^{\dagger}\right)^2 \psi_{\text{R}, j-1, +}
\sim e^{ i \left (\tilde{\phi}_{\textrm{R},j+1}^{-} + 2 \tilde{\phi}_{\textrm{R},j}^{+}+ 2 
\tilde{\phi}_{\textrm{R},j}^{-}+ \tilde{\phi}_{\textrm{R}, j-1 }^{+} \right )}, \nonumber \\ 
O_{\textrm{L}, j} (x) &= \psi_{\textrm{L}, j+1, +} \left ( \psi_{\textrm{L}, j, -}^{\dagger}\right)^2 \left ( \psi_{\textrm{L}, j, +} \right)^2 \psi_{\textrm{L}, j-1, -}^{\dagger}
\sim e^{ i \left (\tilde{\phi}_{\textrm{L},j-1}^{-} + 2 \tilde{\phi}_{\textrm{L},j}^{+}+ 2 
\tilde{\phi}_{\textrm{L},j}^{-}+ \tilde{\phi}_{\textrm{L}, j+1 }^{+} \right )},
\end{align}
with $2 \leq j \leq 2 N -1$.
Note that the physical processes corresponding to those operators conserve the total momentum so that the fast oscillation terms in Eq.~\eqref{supeq:fermionicoperators} are cancelled with each other and hence do not appear in Eq.~\eqref{sup:threeparticleprocess}. 
We next consider a pair of wires of $j = 2k - 1$ and $j = 2k$ ($1 \leq k \leq N$) and define, for every $\eta$ and $k$, two chiral bosonic fields as 
\begin{align} \label{supeq:chiralfields12}
\phi_{\textrm{R}, k, 1}^{\eta} \equiv \tilde{\phi}_{\textrm{R}, 2k-1}^{\eta} + (1 + \eta) ( \tilde{\phi}_{\textrm{R}, 2k}^{+}+  \tilde{\phi}_{\textrm{R}, 2k}^{-} ),& \quad \phi_{\textrm{R}, k, 2}^{\eta} \equiv \tilde{\phi}_{\textrm{R}, 2k}^{\eta} + (1-\eta)( \tilde{\phi}_{\textrm{R}, 2k-1}^{+}+ \tilde{\phi}_{\textrm{R}, 2k-1}^{-} )\,, \nonumber \\\phi_{\textrm{L}, k, 1}^{\eta} \equiv \tilde{\phi}_{\textrm{L}, 2k}^{\eta} + (1 + \eta)( \tilde{\phi}_{\textrm{L}, 2k-1}^{+}+ \tilde{\phi}_{\textrm{L}, 2k-1}^{-} ),& \quad \,
\phi_{\textrm{L}, k, 2}^{\eta} \equiv \tilde{\phi}_{\textrm{L}, 2k-1}^{\eta} + (1 - \eta) ( \tilde{\phi}_{\textrm{L}, 2k}^{+}+  \tilde{\phi}_{\textrm{L}, 2k}^{-} ). 
\end{align}
In terms of the new chiral fields, the gap-forming operators Eq.~\eqref{sup:threeparticleprocess} are written as
\begin{align} \label{supeq:bulkgap}
    O_{\textrm{R}, 2k - 1} \sim  e^{i ( \phi_{\textrm{R}, k-1, 2}^{+} + 
    \phi_{\textrm{R}, k, 2}^{-})}\,,
    &\quad
    O_{\textrm{R}, 2k} \sim  e^{i ( \phi_{\textrm{R}, k, 1}^{+} + 
    \phi_{\textrm{R}, k+1, 1}^{-})}\,, \nonumber \\
     O_{\textrm{L}, 2k-1} \sim  e^{i ( \phi_{\textrm{L}, k-1, 1}^{-} + 
    \phi_{\textrm{L}, k, 1}^{+})}\,,&\quad
    O_{\textrm{L}, 2k} \sim  e^{i ( \phi_{\textrm{L}, k, 2}^{-} + 
    \phi_{\textrm{L}, k+1, 2}^{+})} \,.
\end{align}
From the commutation relations of the original fields (Eq.~\eqref{supeq:commutationrelationoriginal}), one can derive the commutation relations of the new fields 
\begin{align} \label{supeq:Kmatrix}
    \left[\partial_x \phi_{r, k, \ell}^{\eta} (x), \phi_{r', k', \ell'}^{\eta'} (x') \right ] = 2 i \eta \pi \delta_{r r'} \delta_{\eta \eta'} \delta_{k k'} K_{\ell \ell'} \delta (x - x'), \quad \text{with}\,\, K = \begin{pmatrix} 1 & 2 \\ 2 & 1 \end{pmatrix}. 
\end{align}
Furthermore, expressing the density operator of a pair of the wires in terms of the new fields
\begin{align} \label{supeq:tvector}
\frac{1}{2 \pi}\sum_{\eta} \partial_x (\tilde{\phi}_{r, 2k-1}^\eta+ \tilde{\phi}_{r, 2k}^\eta)  
= \sum_{\ell \ell'} \sum_{\eta} t_\ell K^{-1}_{\ell \ell'} \big (\frac{\partial_x \phi_{r, k, \ell'}^{\eta}}{2 \pi} \big), 
\end{align}
one can identify the charge vector $\vec{t}= (1, 1)^T$.  The obtained $K$ matrix and the charge vector $\vec{t}$ in Eqs.~\eqref{supeq:Kmatrix} and \eqref{supeq:tvector} are identical to those for a $\nu = 2/3$ hierarchical state~\cite{Wen1995, Teo2014}. Such a $K$ matrix and a charge vector $\vec{t}$ describe essential properties of a fractional quantum Hall state, fully characterizing its own topological order. In particular, a generic excitation takes the form $T_{\vec{n}} =e^{i \vec{n}^T \cdot \vec{\bar{\vec{\phi}}}_{r,k}^\eta} $ with 
\begin{align} \label{supeq:transformationbosonic}
    \vec{\bar{\vec{\phi}}}_{r,k}^\eta = K^{-1} \vec{\phi}_{r,k}^\eta\,,
\end{align}
where the integer-valued vector $\vec{n}$ describes how many of each type of quasiparticle is created or annihilated, and  
$\vec{\phi}_{r,k}^\eta = (\phi_{r,k,1}^\eta, \phi_{r,k,2}^\eta)$. This excitation has a charge $Q= \vec{t}^T K^{-1} \vec{n}$ and an exchange statistics angle $\theta = \pi \vec{n}^T K^{-1} \vec{n}$~\cite{Wen1995}. 


The $K$ matrix, Eq.~\eqref{supeq:Kmatrix}, is diagonalized by using different bases of bosonic fields. Technically, this diagnonalization is achieved by the transformation $\bar{\phi} \rightarrow W \bar{\phi}$ with a matrix $W \in SL(2, \mathbb{Z})$ that belongs to the special linear group; $W$ is a $2 \times 2$ matrix with integer-values matrix elements and the unit determinant. Such a diagonalization procedure results in the transformation of the $K$ matrix and the $\vec{t}$ vector~\cite{Wen1995, Moore1998} as
\begin{align} 
K' = (W^{-1})^{T} K W^{-1}, \quad \vec{t}' = (W^{-1})^{T}  \vec{t}\,.
\end{align}
Two different bases of the bosonic fields are often used in the existing literature to describe edge modes of the $\nu = 2/3$ state. 
(i) $\phi_1$ and $\phi_{1/3}$ with  
\begin{align}
    W_1= \begin{pmatrix} 2 & 1 \\ -1 & 0 \end{pmatrix}, \quad 
    K_1 = \begin{pmatrix} 1 & 0 \\ 0 & -3 \end{pmatrix}, \quad
    t_1=  \begin{pmatrix} 1 \\ 1 \end{pmatrix}. 
\end{align}
The resulting $K$ matrix has the diagonal elements with the opposite sign, which implies that $\phi_1$ and $\phi_{1/3}$ propagate in the opposite directions to each other~\cite{MacDonald1990, Kane1994}. In this basis, an excitation is written as $e^{i \vec{n}_1^T \cdot \vec{\phi}_1}$ with integers $\vec{n}_1 = (n_1, n_{1/3})$ and $\vec{\phi}_1 = (\phi_1, \phi_{1/3})$. This excitation has 
a charge $Q = \vec{n}_1^T (K_1)^{-1} \vec{t}_1 = (n_1 - \frac{n_{1/3}}{3})$ and an exchange statistics angle $\pi \vec{n}_1^T (K_1)^{-1} \vec{n}_1$. 
(ii) Charge-neutral basis $\phi_c$ and $\phi_n$ with 
\begin{align} \label{supeq:chargeneutralbasis}
    W_2= \begin{pmatrix} 1/2 & 1/2 \\ 1/2 & -1/2 \end{pmatrix}, \quad 
    K_2 = \begin{pmatrix} 6 & 0 \\ 0 & -2 \end{pmatrix}, \quad
    t_2=  \begin{pmatrix} 2 \\ 0 \end{pmatrix}. 
\end{align}
In this basis, generic excitations take the form  
\begin{align}
    T_{\vec{n}_2} = e^{i (\phi_c n_c + \phi_n n_n)}. 
\end{align}
with the integer-valued vector $\vec{n}_2 = (n_c, n_n)$ and have a charge  $Q = \vec{n}_2^T (K_2)^{-1} \vec{t}_2 = n_c/3$ and statistics $\theta = \pi \vec{n}_2^T (K_2)^{-1} \vec{n}_2 = \frac{\pi}{2} (n_c^2/3 - n_n^2)$. It is important to note that the transformation matrix of Eq.~\eqref{supeq:chargeneutralbasis} does not belong to $SL(2, \mathbb{Z})$. As a consequence,  $n_c$ and $n_n$ are not independent of each other and must obey $n_c + n_n \in \mathbb{Z}_{\text{even}}$~\cite{Naud2000}. For instance, the $e/3$-charged excitations ($n_c = 1$) always accompany neutral excitations. This charge-neutral basis looks unnatural due to this constraint, but it provides a convenient way to describe the physics of the present study and thus 
we will use this charge-neutral basis throughout this paper. 

Employing Eqs.~\eqref{supeq:transformationbosonic} and \eqref{supeq:chargeneutralbasis}, we define charge and neutral modes for each individual wire
as
\begin{align} \label{supeq:chargeneutralmodes}
\phi_{c, r, k}^{\eta} \equiv \frac{1}{6} (\phi_{r, k, 1}^{\eta} + \phi_{r, k, 2}^{\eta}), \quad
\phi_{n, r, k}^{\eta} \equiv \frac{1}{2} (\phi_{r, k, 1}^{-\eta} - \phi_{r, k, 2}^{-\eta}). 
\end{align}
From Eq.~\eqref{supeq:Kmatrix}, one can show the following commutation relation for those charge and neutral modes
\begin{align} \label{supeq:commutationchargeneutral}
[\phi_{\ell, r, k}^{\eta}(x), \phi_{\ell', r', k'}^{\eta'} (x')] = i \pi  \eta |(K_2^{-1})_{\ell \ell'}| \delta_{\ell \ell'} \delta_{k k'} \delta_{\eta \eta'} \text{sgn} (x - x')\,,
\end{align} 
for $\ell = c, n$. The gap-forming interactions Eq.~\eqref{supeq:bulkgap} are also written in this charge-neutral basis  
\begin{align} \label{supeq:bulkgap2}
    O_{\textrm{R}, 2k +1} &\sim   e^{3i (\phi_{\textrm{c},\textrm{R}, k}^{+} + \phi_{\textrm{c},\textrm{R}, k+1}^{-})} e^{- i  (\phi_{\textrm{n},\textrm{R}, k}^{-} + \phi_{\textrm{n},\textrm{R}, k+1}^{+})},
    \nonumber \\ 
    O_{\textrm{R}, 2k} &\sim  e^{3i  (\phi_{\textrm{c},\textrm{R}, k}^{+} + \phi_{\textrm{c},\textrm{R}, k+1}^{-})}
    e^{i  (\phi_{\textrm{n},\textrm{R}, k}^{-}+ \phi_{\textrm{n},\textrm{R}, k+1}^{+})}, \nonumber \\
     O_{\textrm{L}, 2k+ 1} &\sim  e^{3i  (\phi_{\textrm{c},\textrm{L}, k}^{-} + \phi_{\textrm{c},\textrm{L}, k+1}^{+})} e^{i  (\phi_{\textrm{n},\textrm{L}, k}^{+} + \phi_{\textrm{n},\textrm{L}, k+1}^{-})},
    \nonumber \\ 
    O_{\textrm{L}, 2k} &\sim  e^{3 i  (\phi_{\textrm{c},\textrm{L}, k}^{-} + \phi_{\textrm{c},\textrm{L}, k+1}^{+})}
    e^{-i  (\phi_{\textrm{n},\textrm{L}, k}^{+}+ \phi_{\textrm{n},\textrm{L}, k+1}^{-})}.
\end{align}
Collecting all terms together, the Hamiltonian associated with the three-particle involving processes is written as 
\begin{align} \label{supeq:bulkHamiltonian}
    H_{\text{bulk}} = - \sum_{r = \text{R}/\text{L} = \pm 1} \sum_{k=1}^{N-1}  & \Big ( g_{r, 2k+1} \cos \Big ((\phi_{\textrm{c},r, k}^{r} + \phi_{\textrm{c},r, k+1}^{-r}) - r (\phi_{\textrm{n},r, k}^{-r} + \phi_{\textrm{n},r, k+1}^{r}) \Big ) \nonumber \\ & \quad
    +g_{r, 2k} \cos \Big ( (\phi_{\textrm{c},r, k}^{r} + \phi_{\textrm{c},r, k+1}^{-r}) + r(\phi_{\textrm{n},r, k}^{-r} + \phi_{\textrm{n},r, k+1}^{r}) \Big ) 
\Big ), 
\end{align}
with $r = \text{R}/ \text{L} = \pm 1$. Here $g_{r, k}$ denotes the strength of the three-particle processes. 
Under the assumption that all of those gap forming operators are relevant in the renormalization group, all $g_{r, k}$'s flow to the strong coupling regime where  
all the cosine terms find their own minimum, and the fluctuations around the minimum are massive. Since the charge and neutral modes commute with each other, i.e., $[\phi_{c, r, k}^{\eta}(x), \phi_{n, r', k'}^{\eta'} (x')] =0$, the charge and neutral modes in the bulk can be separately gapped out. The four boundary modes ($\phi_{c, \text{R}, k=1}^{-} \equiv \phi_{c}^{-}$,  $\phi_{n, \text{R}, k=1}^{+} \equiv \phi_{n}^{+}$, $\phi_{c, \text{L}, k=1}^{+} \equiv \phi_{c}^{+}$, $\phi_{n, \text{L}, k=1}^{-} \equiv \phi_{n}^{-}$) are, on the other hand, decoupled in Eq.~\eqref{supeq:bulkHamiltonian} and thus they remain gapless. From Eq.~\eqref{supeq:commutationchargeneutral}, one can show that the boundary modes fulfill the following commutation relation
\begin{align} \label{supeq:commutationrelationboundary}
[\phi_{\ell}^{\eta}(x), \phi_{\ell'}^{\eta'} (x')] = i \pi  \eta (K^{-1}_{0})_{\ell \ell'}\delta_{\eta \eta'} \text{sgn} (x - x')\,,
\end{align}
with $\ell, \ell' = n, c$ and $K_0=\begin{pmatrix}
   2 & 0 \\
  0 & 6 \\
   \end{pmatrix}$. Such two counter-propagating boundary modes at each quantum Hall puddle
is largely consistent with recent experiments where upstream modes are detected from a shot noise measurement or thermal transport~\cite{Banerjee2017, Melcer2022, Kumar2022, Srivastav2022}. 

\section{Interaction parameters for boundary modes from a microscopic Hamiltonian}
In this section, we find an effective Hamiltonian for the boundary modes that are left behind after gapping out the bulk modes as discussed in Sec.~\ref{supsec:wireconstruction}. To do so, we use a simple microscopic model for the wires and derive interaction parameters of the boundary modes from the parameters of the microscopic model. 

We consider a microscopic Hamiltonian, given by 
\begin{align} \label{supeq:wireinteraction}
    H_{\rm{wire}} =  \int dx \big (\sum_{r=R,L} \big ( -i \hbar v_F \sum_{j=1}^{2N} \sum_{\eta = \pm} \eta \psi_{r, j, \eta}^{\dagger} \partial_x \psi_{r, j, \eta}
    + \frac{U}{2} \sum_{j=1}^{2N} \rho_{r, j}^2  
   + V \sum_{j=1}^{2N-1} \rho_{r, j}\rho_{r, j+1}\big )  + V' \rho_{L, 1} \rho_{R,1} \big) + H_{\text{bulk}}, 
\end{align}
with the long-wavelength electron density $\rho_{r, j} (x) \equiv \sum_{\eta = \pm } \psi_{r,j,\eta}^{\dagger} (x) \psi_{r,j,\eta} (x)$ of wire $j$ in QH puddle $r$. 
Here $v_F$ is bare Fermi velocity of the wires, $U$ the intra-wire interaction, $V$ the nearest neighbor intra-puddle interaction, and $V'$ the nearest neighbor inter-puddle interaction. Each term of Eq.~\eqref{supeq:wireinteraction} preserves the total momentum and thus it survives after the integration in $x$. The last term $H_{\text{bulk}}$ represents the gap-forming interactions \eqref{sup:threeparticleprocess}. Here, we neglect possible back-scattering terms to change the propagating direction of the modes since those terms compete with the gap-forming interactions $H_{\text{bulk}}$ and have to be suppressed in order to develop a QH state.

We next bosonize Eq.~\eqref{supeq:wireinteraction} and write it in terms of charge and neutral modes. We assume that $H_{\text{bulk}}$ is relevant in the renormalization group so that the bulk modes are gapped out and they do not contribute to the low-energy dynamics. We instead keep the gapless modes $\phi_{c}^{\pm}$ and $\phi_{n}^{\pm}$ at the interface between the QH puddles.
In the basis of $\vec{\phi} = (\phi_c^+,\phi_c^-, \phi_n^+,\phi_n^-)$, the resulting action for the boundary modes takes the form 
\begin{align} \label{supeq:forward}
S_{0} = & - \frac{1}{4\pi} \int dx dt \left[\left (\partial_x \vec{\phi}^T \right )\cdot ( K \cdot \partial_t \vec{\phi} + V_{\text{int}} \cdot \partial_x \vec{\phi})  \right]\,.
\end{align} 
Here $K=\begin{pmatrix}
   6 \sigma_z & 0 \\
   0 &   2\sigma_z \\
    \end{pmatrix}$, while the interaction matrix $V_{\text{int}}$ is
\begin{align}
V_{\text{int}} =\begin{pmatrix} \label{supeq:interactionmatrix}
v_c & g_c & v_{cn}^{++} & v_{cn}^{+-} \\ 
g_c & v_c & v_{cn}^{-+} & v_{cn}^{--} \\ 
v_{cn}^{++} & v_{cn}^{-+} & v_n & g_n \\ 
v_{cn}^{+-} & v_{cn}^{--} & g_n & v_n
\end{pmatrix},
\end{align}
with 
\begin{align} \label{supeq:interactioninwireparams}
v_c &= \frac{U+V + 14 \pi v_F}{\pi}, \quad v_n = \frac{U-V+6 \pi v_F}{\pi}, \quad v_{cn}^{\pm \mp} = \pm 8v_F, 
\nonumber \\ 
g_c &= \frac{V'}{2 \pi}, \quad g_n = - \frac{V'}{2 \pi}, \quad v_{cn}^{\pm \pm} = \pm \frac{V'}{2 \pi}.
\end{align}
Here we neglect further renormalizations ($\sim 1/E_{\text{gap}}$) of the interaction terms originating from integrating out the bulk modes
since the effect of the renormalizations becomes negligible with a sufficiently large bulk gap $E_{\text{gap}}$.

Note that with repulsive interactions $V'>0$, the sign of $g_n$ is negative and hence $\phi_n^+$ and $\phi_n^-$ attract each other. 
This attractive interaction can be understood by the Pauli exclusion principle. For the spin-polarized QH states discussed here, the spatial part of the wave function is anti-symmetric such that a particle is surrounded by an hole to minimize the exchange energy. As a consequence, the interaction between the neighboring sites is attractive. In the hole-conjugated states, such an anti-symmetric feature of the wave function is encoded on neutral modes that govern the spatial modulation of charge-neutral density fluctuations. Therefore, the neighboring neutral modes may attract each other. Although our finding relies on the wire-construction approach, we expect $g_n <0$ is a generic feature for any hole-conjugated states since it is attributed to a fundamental fermionic nature of the wave function, the Pauli exclusion principle. 



\section{Phase diagram}
In this section, we investigate a zero temperature phase diagram for the boundary modes ($\phi_c^{\pm}$ and $\phi_n^{\pm}$). To do so, we introduce some operators that couple the boundary modes and find the most relevant operator to govern the low-energy physics in a renormalization group manner. Specifically, we calculate the scaling dimension of the operators with respect to the action \eqref{supeq:forward} and find the operator with the smallest scaling dimension. The phase diagram is obtained in terms of the parameters of wires \eqref{supeq:wireinteraction}. 

\begin{figure}
\includegraphics[width=0.95\columnwidth]{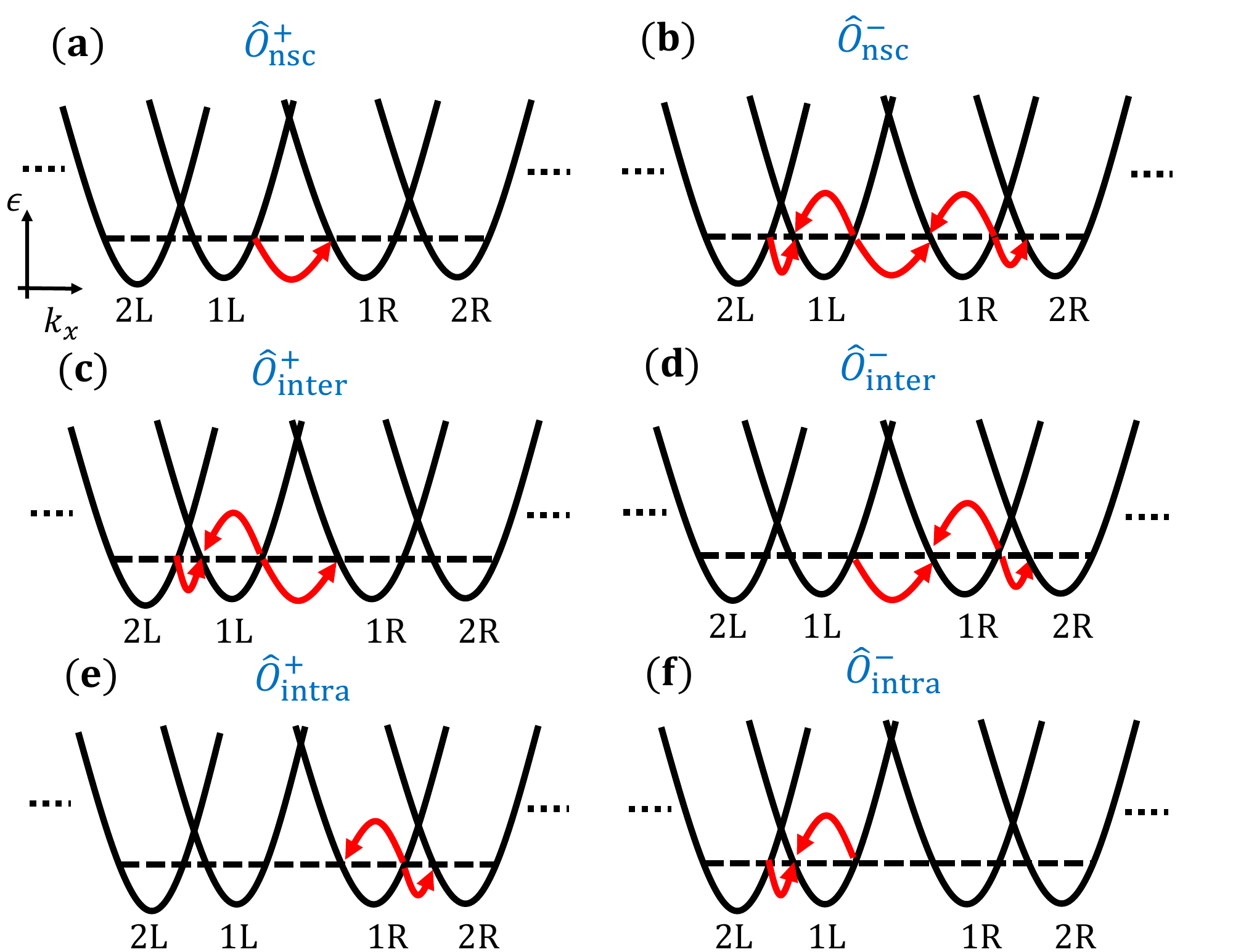} 
\caption{{\bf Physical processes} to depict the operators in Eq.~\eqref{supeq:scatteringterm} at the boundary between two quantum Hall puddles. Every process does not preserve the momentum conservation and thus disorder is necessary to facilitate such processes. {\bf(a-b)} Processes to depict the neutral superconductivity terms. Panel (a) is equivalent to Fig.~2(d) in the main text. {\bf(c-d)} Processes to depict the backscattering terms of the neutral constituent, $\hat{O}_{\textrm{inter}}^{\pm}$. The processes in panels (a)-(d) involve electron tunneling between the two quantum Hall puddles. By contrast, the processes displayed in panels {\bf(e-f)} involve electron tunneling within each puddle. 
}
\label{supfig:wires}
\end{figure}

We begin our discussion with the introduction of the following operators for the boundary modes
\begin{align} \label{supeq:intermode-scattering}
   S_{\rm{inter}} = \int dx dt \left (\xi_{\textrm{nsc}}^{\pm} \hat{O}_{\textrm{nsc}}^{\pm} + \xi_{\textrm{inter}}^{\pm} 
   \hat{O}_{\textrm{inter}}^{\pm} +  \xi_{\textrm{intra}}^{\pm} \hat{O}_{\textrm{intra}}^{\pm} + h.c.\right ).
\end{align}
with
\begin{align} \label{supeq:scatteringterm}
\hat{O}_{\textrm{nsc}}^{+} &= e^{ i(\phi_n^{+} - \phi_{n}^{-})} e^{3 i (\phi_c^{+} + \phi_{c}^{-})}
= e^{i (\tilde{\phi}_{\text{R}, j=1}^{\eta = -} + \tilde{\phi}_{\text{L}, j=1}^{\eta = + })}  \,, \nonumber \\
\hat{O}_{\textrm{nsc}}^{-} &= e^{- i(\phi_n^{+} - \phi_{n}^{-})} e^{3 i (\phi_c^{+} + \phi_{c}^{-})}
= e^{i (2 \tilde{\phi}_{\text{R}, j=1}^{\eta = -} + 2 \tilde{\phi}_{\text{R}, j=1}^{\eta = +} + \tilde{\phi}_{\text{R}, j=2}^{\eta = -} )} e^{i (2 \tilde{\phi}_{\text{L}, j=1}^{\eta = -} + 2 \tilde{\phi}_{\text{L}, j=1}^{\eta = +} + \tilde{\phi}_{\text{L}, j=2}^{\eta = +} )}  \,, \nonumber \\
\hat{O}_{\textrm{inter}}^{+} &= e^{i(\phi_n^{+} + \phi_{n}^{-})} e^{3 i (\phi_c^{+} + \phi_{c}^{-})}
= e^{i \tilde{\phi}_{\text{R}, j=1}^{\eta = -} } e^{i (2 \tilde{\phi}_{\text{L}, j=1}^{\eta = -} + 2 \tilde{\phi}_{\text{L}, j=1}^{\eta = +} + \tilde{\phi}_{\text{L}, j=2}^{\eta = +} )} \,,\nonumber \\\hat{O}_{\textrm{inter}}^{-} &= e^{ - i(\phi_n^{+} + \phi_{n}^{-})} e^{3 i (\phi_c^{+} + \phi_{c}^{-})}
=  e^{i (2 \tilde{\phi}_{\text{R}, j=1}^{\eta = -} + 2 \tilde{\phi}_{\text{R}, j=1}^{\eta = +} + \tilde{\phi}_{\text{R}, j=2}^{\eta = -} )} e^{i \tilde{\phi}_{\text{L}, j=1}^{\eta = +} } \,, \nonumber \\
\hat{O}_{\textrm{intra}}^{\pm} &=  e^{2i  \phi_n^{\pm}} =e^{ \mp i (\tilde{\phi}_{\text{R}/\text{L}, j=1}^{\eta = \mp} + 2 \tilde{\phi}_{\text{R}/\text{L}, j=1}^{\eta = \pm} + \tilde{\phi}_{\text{R}/\text{L}, j=2}^{\eta = \mp} )}\,.
\end{align}
Note that according to the Haldane topological stability condition~\cite{Haldane1995}, each individual operator, $\hat{O}_{\text{nsc}}^{\pm}$ and $\hat{O}_{\text{inter}}^{\pm}$, can open a mobility gap, leading to the localization of boundary modes. Furthermore, neutral superconductivity operators $\hat{O}_{\text{nsc}}^{\pm}$ and inter-puddle electron-tunneling operators $\hat{O}_{\text{inter}}^{\pm}$ compete with each other and thus they are not accommodated simultaneously. What operators win against the others at low-energy are determined by renormalization group equations, which will be shown below. The right most expression of each operator in Eq.~\eqref{supeq:scatteringterm} is written in terms of the wire basis $\tilde{\phi}_{r, j}^{\eta}$ (see Eq.~\eqref{supeq:fermionicoperators}), which describes the modes with chirality $\eta = \pm$ in the 
the $j$th wire of puddle $r=\text{R/L}$.

Figure \ref{supfig:wires} depicts the physical process corresponding to each individual operator in Eq.~\eqref{supeq:scatteringterm}. 
Note that every process does not preserve the momentum conservation. When the inter-wire distance $a$ equals to the inter-puddle distance $b$  (see Fig.~2(b) in the main text for the configuration), the $\hat{O}_{\textrm{inter}}^{\pm}$ operators (depicted in Figs.~\ref{supfig:wires}(c) and (d)) are identical to the bulk gap-forming interaction term, Eq.~(3) and Fig.~2(c) in the main text, where the momentum is instead conserved. By contrast, the processes associated with $\hat{O}_{\textrm{inter}}^{\pm}$ for generic $b \neq a$ no longer satisfy the momentum conservation. The fact that the operators displayed in Eq.~\eqref{supeq:scatteringterm} break the momentum conservation indicates that disorder is crucial for the emergence of such operators. 

\begin{figure}
\includegraphics[width=0.95\columnwidth]{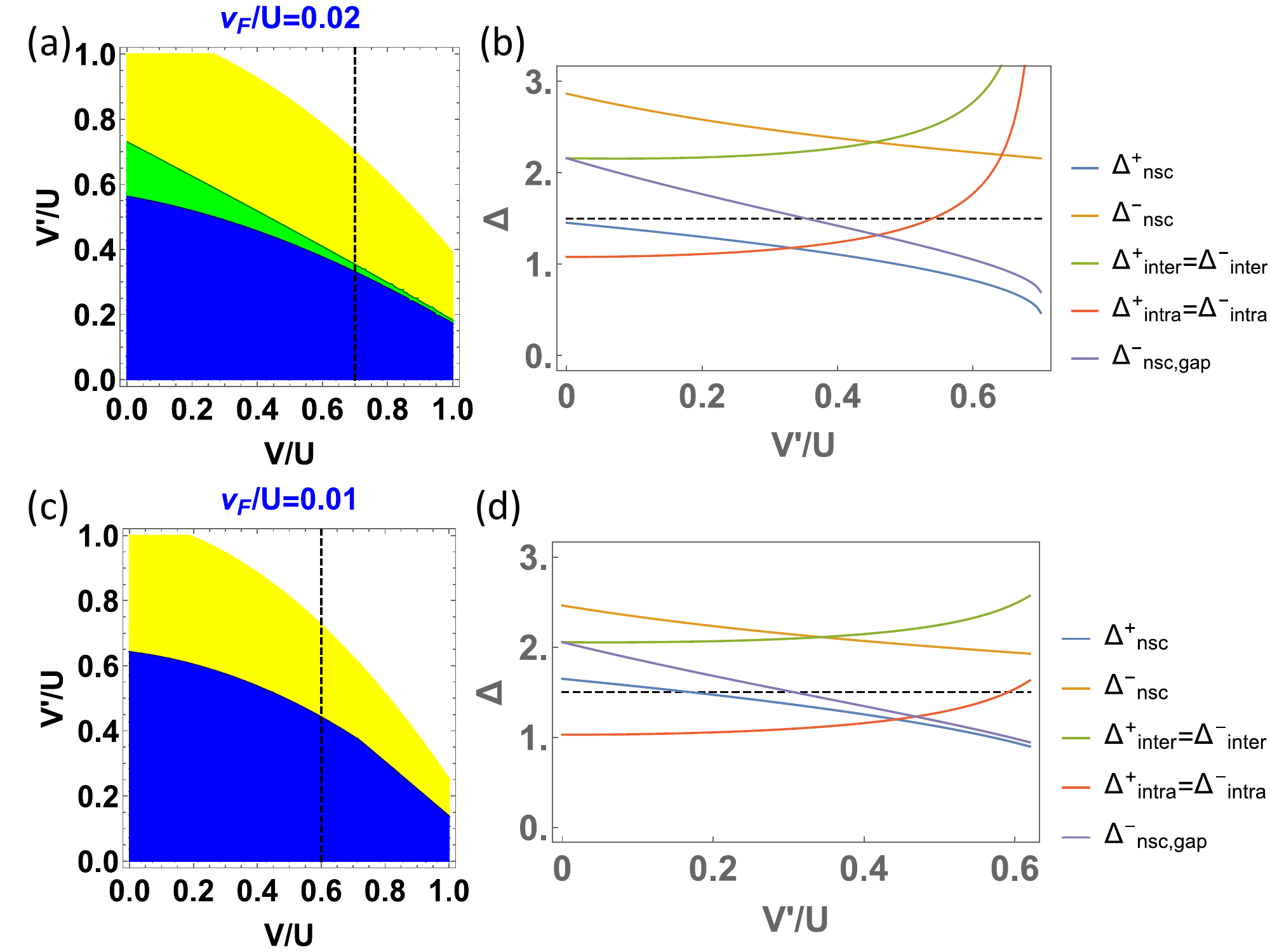} 
\caption{{\bf Zero-temperature phase diagrams} for the boundary modes in the parameter space $(V/U, V'/U)$ of wires, see Eq.~\eqref{supeq:wireinteraction} for more details on the parameters. {\bf (a, c)} The phase diagrams are displayed with two different $v_F$: (a) $v_F = 0.02 U$ and (c) $v_F = 0.01 U$. Three distinct phases appear; (i) a phase with four gapless modes (denoted in blue), (ii) a phase with two gapless and two gapped modes (green), and a fully gapped phase (yellow). In the white region, the boundary modes are no longer stable where the $V_\text{int}$-matrix \eqref{supeq:interactionmatrix} is not positive-definite. 
In particular, the fully gapped phase stabilizes the neutral superconductivity where both of the $\hat{O}_{\text{nsc}}^{\pm}$ operators are relevant (i.e., $\Delta_{\text{nsc}}^{\pm} < 3/2$) with the smallest scaling dimensions than the other operators in Eq.~\eqref{supeq:scatteringterm}. Panel {\bf (b)} and {\bf (d)} plots the scaling dimensions of the operators Eq.~\eqref{supeq:scatteringterm} as a function of $V'/U$ along the black dashed line in panel (a) and (c), respectively. While the intra-puddle electron tunneling terms $\hat{O}_{\text{intra}}^{\pm}$ have the smallest scaling dimension in the blue region, the yellow region is characterized by the $\hat{O}_{\text{nsc}}^{\pm}$ terms, see the text for more details.  
}
\label{supfig:phasediagram}
\end{figure}

For weak disorder, the fate of the operators \eqref{supeq:scatteringterm} is determined by the renormalization group equations~\cite{Giamarchi1988}, 
\begin{align} \label{subeq:renormaliztiongroup}
    \frac{d W_{\ell}^{\pm}}{d \ln \mathcal{L}} = (3 - 2 \Delta_{\ell}^{\pm} ) W_{\ell}^{\pm}\,. 
\end{align}
Here $\Delta_{\ell}^{\pm}$ is the bare scaling dimension of the operator $O_{\ell}^{\pm}$ for $\ell= (\textrm{nsc}, \textrm{inter}, \textrm{intra})$ which is evaluated at the action \eqref{supeq:forward}.  
The coupling strengths  $\xi_{\ell}^{\pm}$ in Eq.~\eqref{supeq:intermode-scattering} are assumed to follow a Gaussian distribution and thus satisfy 
$\langle \xi^{\pm}_{\ell} (x) (\xi^{\pm}_{\ell'})^* (x') \rangle_{\text{dis}} = W_{\ell}^{\pm} \delta_{\ell \ell'} \delta (x - x')$. 
We further assume that the disorder is sufficiently weak such that we only keep the leading order in $W_{\ell}^{\pm}$ in Eq.~\eqref{subeq:renormaliztiongroup}. 
The renormalization of $\Delta_{\ell}^{\pm}$ is also neglected. 

Equation \eqref{subeq:renormaliztiongroup} indicates that the operator with the smallest $\Delta_{\ell}^{\pm}$ governs the low-energy physics. The disorder strength $W_{\ell}^{\pm}$ of the operator with the smallest $\Delta_{\ell}^{\pm}$ grows more rapidly and when $\Delta_{\ell}^{\pm} < 3/2$, it may reach a strong coupling regime earlier than the other operators. Once an operator $\hat{O}_{\ell}^{\pm}$ arrives in a strong coupling regime, the operators with different $\ell' \neq \ell$ are highly fluctuating, and thus they become strongly suppressed. 
Therefore, the operators $\hat{O}_{\ell}^{\pm}$ with the smallest $\Delta_{\ell}^{\pm}$ will survive in the competition with the other operators.

With this spirit in mind, we compute the scaling dimensions of the operators in Eq.~\eqref{supeq:intermode-scattering} under the action Eq.~\eqref{supeq:forward} with the parameters Eq.~\eqref{supeq:interactioninwireparams} derived from the Hamiltonian Eqs.~\eqref{supeq:wireinteraction}. Among those operators, we identify the operator that is renormalization-group relevant (i.e., has the smallest scaling dimension). To do so, we diagonalize Eq.~\eqref{supeq:forward}
by the transformation of $\vec{\phi} = M_2 M_1  \overline{\vec{\phi}}$ with the matrices $M_1 \in SO(2,2)$ and $M_2$ satisfying $M_1^T (\mathbb{1}_2 {\otimes} \sigma_z) M_1 = (\mathbb{1}_2 {\otimes} \sigma_z)$ and $M_2^T K M_2 = \mathbb{1}_2 {\otimes} \sigma_z$. Then, the scaling dimension $\Delta_{\vec{m}}$ for generic operators $e^{i \vec{m} \cdot \vec\phi}$ is given by the following formula~\cite{Moore1998}
\begin{align}
    \Delta_{\vec{m}} = \frac{1}{2} \vec{m}^T M_2 M_1 M_1^T M_2^T  \vec{m}\,.
\end{align}
By identifying the operator with the smallest scaling dimension, we draw phase diagrams for the boundary modes in Figs.~\ref{supfig:phasediagram}(a) and (c) in the parameter space $(V/U, V'/U)$. The phase diagrams are characterized by three distinct phases: (i) a phase with four gapless modes (blue regions), (ii) a partially gapped phase (green), and (iii) a fully gapped phase (yellow). 

In the gapless phase (the blue regions in Figs.~\ref{supfig:phasediagram}(a) and (c)), the most relevant operators are $\hat{O}_{\textrm{intra}}^{\pm}$. Those operators were investigated before in Refs.~\cite{Kane1994,Kane1995} to describe the equilibration in an edge of the $\nu = 2/3$ quantum Hall state. Following the Haldane's criterion~\cite{Haldane1995} for the topological stability of edge modes against localization, it is shown that those operators can not generate a gap; One can show that $\vec{m}^T K_2^{-1} \vec{m} \neq 0$ with $\vec{m}^T = (0,2)$ and $K_2$ (Eq.~\eqref{supeq:chargeneutralbasis}) in the charge-neutral basis (see also Refs.~\cite{Kao1999, Spanslatt2023} for examples of a phase transition of topologically unstable states). Moreover, when $\hat{O}_{\textrm{intra}}^{\pm}$ become relevant, the boundary modes of each puddle flow to a disorder-dominated fixed point where the charge modes $\phi_c^\pm$ are completely decoupled with the neutral modes $\phi_n^\pm$~\cite{Kane1994, Kane1995}. At this fixed point, one can show that the operators $\hat{O}_{\textrm{inter}}^{\pm}$ and $\hat{O}_{\textrm{nsc}}^{\pm}$ have the identical scaling dimensions. From this, we conclude that neither $\hat{O}_{\textrm{inter}}^{\pm}$ nor $\hat{O}_{\textrm{nsc}}^{\pm}$ is stabilized at the fixed point and thus may not generate a mobility gap. Therefore, all boundary modes remain gapless even though the $\hat{O}_{\textrm{intra}}^{\pm}$ terms stabilize at low temperatures. This phase is fully characterized by the central charge $c = 2$, which reflects the two gapless modes in both directions. As shown in Fig.~\ref{supfig:phasediagram}, this phase is achieved for a relatively small $V$ and $V'$.

The scaling dimensions of each operator in Eq.~\eqref{supeq:scatteringterm} along the dashed lines of Figs.~\ref{supfig:phasediagram}(a) and (c) are plotted in Figs.~\ref{supfig:phasediagram}(b) and (d), respectively. Upon increasing $V'>0$, the scaling dimensions $\Delta^{\pm}_{\text{nsc}}$ of the neutral superconductivity terms $\hat{O}_{\textrm{nsc}}^{\pm}$ decrease, see the blue and orange lines in Figs.~\ref{supfig:phasediagram}(b) and (d). It indicates that the neutral superconductivity terms become more stabilized for a stronger repulsive interaction $V'>0$ since the $V'$-interaction in Eq.~\eqref{supeq:wireinteraction} induces an attractive interaction between $\phi_n^+$ and $\phi_n^-$. Note  the negative sign of $g_n$ in Eq.~\eqref{supeq:interactioninwireparams}. In particular, $\Delta^{+}_{\text{nsc}}$  even falls below $\Delta^{\pm}_{\text{intra}}$ at certain $V'$, rendering $\hat{O}^{+}_{\text{nsc}}$ the most relevant. In this regime, the low-energy physics is governed by the effective action 
\begin{align} \label{supeq:effectiveaction}
    S_{\text{low}} = 2 \int dx dt |\xi_{\text{nsc}}^{+}| \cos \left (\phi_n^+ - \phi_n^- + 3 (\phi_c^+ + \phi_c^-) + \Theta^+ \right) 
\end{align}
with $\xi_{\text{nsc}}^+ = |\xi_{\text{nsc}}^{+}| e^{i \Theta^+} $. Once this cosine term flows to a strong coupling regime, the charge and neutral modes inside the cosine term are pinned to one of the minima of the cosine terms such that $\phi_n^+ - \phi_n^- + 3 (\phi_c^+ + \phi_c^-) + \Theta^+ = 2 \pi n$ with $n \in  \mathbb{Z}$. At the same time, the other terms except for $\hat{O}_{\text{nsc}}^-$ become highly fluctuating and hence vanish. The $\hat{O}_{\text{nsc}}^-$ term, in contrast, survives and even stabilizes 
after the pinning of the phase factor of $\hat{O}_{\text{nsc}}^+$ as $\hat{O}_{\text{nsc}}^-$ commutes with $\hat{O}_{\text{nsc}}^+$. After the pinning, the $\hat{O}_{\text{nsc}}^-$ term reads (up to a constant factor)
\begin{align} \label{eq:nscminustermaftergapping}
    \hat{O}_{\text{nsc}}^- \approx e^{2 i (\phi_n^+ - \phi_n^-)}\,.
\end{align}
The scaling dimension of Eq.~\eqref{eq:nscminustermaftergapping} is also plotted in Figs.~\ref{supfig:phasediagram}(b) and (d). Compared with the scaling dimension $\Delta_{\text{nsc}}^-$ before the pinning (green lines), the scaling dimension after the pinning (purple lines) is significantly reduced, reflecting further stabilization of $\hat{O}_{\text{nsc}}^-$ term.

When the phase factor inside the cosine term in Eq.~\eqref{supeq:effectiveaction} is pinned, two different senarios are possible. First, the scaling dimension of Eq.~\eqref{eq:nscminustermaftergapping} is still above $3/2$ so that $\hat{O}_{\text{nsc}}^-$ does not open a gap. This phase is referred to as {\it partially gapped phase} (denoted as green in Fig.~\ref{supfig:phasediagram}(a)) and it is characterized by the central charge $c=1$. Second, the scaling dimension of Eq.~\eqref{eq:nscminustermaftergapping} is less than $3/2$ and thus the phase factor in Eq.~\eqref{eq:nscminustermaftergapping} can be also pinned at low temperatures. Then, all boundary modes are fully gapped with the zero central charge $c= 0$ (the yellow regions in Figs.~\ref{supfig:phasediagram}(a) and (c)). Those three different phases, (i) the gapless phase (blue), (ii) the partially gapped phase (green), and (ii) the fully gapped phase (yellow) have distinct central charges; phase (i) $c = 2$, (ii) $c= 1$, (iii) $c = 0$. Therefore, each individual phase can be identified by the measurement of thermal conductance. 

On the other hand, the inter-puddle interactions $\hat{O}_{\textrm{inter}}^{\pm}$ always have larger scaling dimensions than the other operators due to the attractive interaction between $\phi_n^+$ and $\phi_n^-$. Therefore, $\hat{O}_{\textrm{inter}}^{\pm}$ are not crucial at low temperatures. 


\end{document}